\documentclass{ws-p8-50x6-00}

\begin{document}

\title{
The most energetic particles in the Universe}

\author{E. Roulet$^a$,
 D. Harari$^b$ and S. Mollerach$^a$.}

\address{$A$) Depto. de F\'\i sica, U. Nac. de La Plata, CC67, 1900,
Argentina}
\address{$B$) Depto. de F\'\i sica, FCEyN, U. de Buenos Aires, 1428,
Argentina.}

\address{(Plenary talk presented by
E. Roulet at COSMO99, Trieste, Oct. 1999)}

\maketitle

\abstracts{Several issues related to the lensing of ultra-high energy
cosmic rays by the Galactic magnetic field are discussed.}

The spectrum of the cosmic rays (CRs) arriving to the Earth has the
overall shape of a leg, with a knee at $10^{16}$~eV beyond which it
becomes steeper and an ankle at $\sim 5\times 10^{18}$~eV where it becomes
harder again up to the highest energies observed of $3\times 10^{20}$~eV.
The region beyond the ankle, i.e. the foot of the spectrum, is attracting
renewed interest nowadays with the deployment of large detectors, such as
HiRes and Auger, which are expected to find an answer to several issues
raised by the observation of such energetic events. These are:

\noindent $i$) The production mechanism giving rise to such enormous
energies
(relativistic Fermi acceleration, production in decays of topological
defects or heavy relics from the big bang, etc.).

\noindent $ii)$ The nature of the primaries, i.e. whether they are
protons, nuclei,
$\gamma$ rays, neutrinos with new interactions or more exotic objects (see
Berezinsky's talk).

\noindent $iii)$ How they manage to propagate from their sources up to us,
since for
instance protons are attenuated by photopion production off the CMB
photons at energies above $5\times 10^{19}$~eV \cite{gzk}, and similarly
nuclei can
photodisintegrate through interactions with CMB and IR photons
\cite{pu76}. Through
these processes `hadronic' CRs with energies above $10^{20}$~eV would lose
their energy in a few dozens of Mpc, leading to the famous GZK cutoff
which was expected, but not observed, in the CR spectrum.

\noindent $iv)$ The location of the sources, i.e. if CRs are produced in
the Galaxy,
if they are extragalactic (e.g. produced in active galaxies) or uniformly
spread through cosmological distances, as in some topological defect
models.

\begin{figure}[t]
\epsfxsize=25pc 
\epsfbox{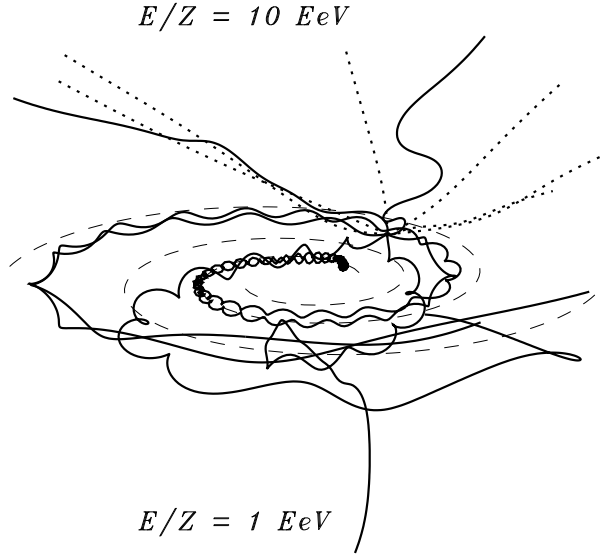} 
\caption{Examples of trajectories of nuclei with $E/Z=1$~EeV
($=10^{18}$~eV) and 10~EeV. At the lower energies particles start to be
trapped by the spiral structure of the regular galactic magnetic field.
  \label{fig:orbits}}
\end{figure}

It is widely believed that below the ankle CRs are protons or nuclei
mostly
of galactic origin. Since the gyroradius of a CR with
energy $E$ and charge $Z$ in a magnetic field $B$
is $R\sim {\rm kpc}(\mu G/B)(E/Z\ 10^{18}{\rm
eV})$, below the ankle the CR trajectories are very curly for the galactic
fields of a few $\mu G$ and one has to describe the propagation in terms
of diffusion and drift. However, above the ankle the gyroradii become
comparable or larger than the scale of the galactic magnetic field, so
that trajectories are straighter and one can start to speak of small CR
deflections due to the magnetic fields. In particular one expects to be
able to roughly trace back the location of the sources, and hence to do
astronomy, with the highest energy events (see fig.~1).

The lack of any obvious observed anisotropy towards the galactic plane
then suggests that in the `foot' of the spectrum CRs are most likely
extragalactic (see however de R\'ujula's talk). In this case, if CRs are
indeed normal hadronic matter (nuclei or protons) the sources should not
be too far (i.e. less than 20-50~Mpc). Correlations of
the observed arrival directions with the location of candidate sources or
with the general direction of the supergalactic plane have been searched
for (but
with no clear evidence of correlations found yet however). When looking
for the source locations it is important to correct for possible magnetic
deflections using plausible models of the Galactic magnetic fields
\cite{st97,ha99}, and eventually also of extragalactic ones if these
were to turn 
out to be very large \cite{bh99}. For instance, fig.~2 shows the 
directions of arrival to the halo  of the highest energy events recorded
by AGASA assuming different CR compositions, from protons up to Fe
nuclei, adopting a bisymmetric spiral model for the galactic magnetic
field \cite{ha99}. Clearly the deflections are sizeable even at these
energies if CRs
are heavy nuclei. To do detailed CR astronomy would require then to know
the
CR composition to some extent.

\begin{figure}[t]
\epsfxsize=30pc 
\epsfbox{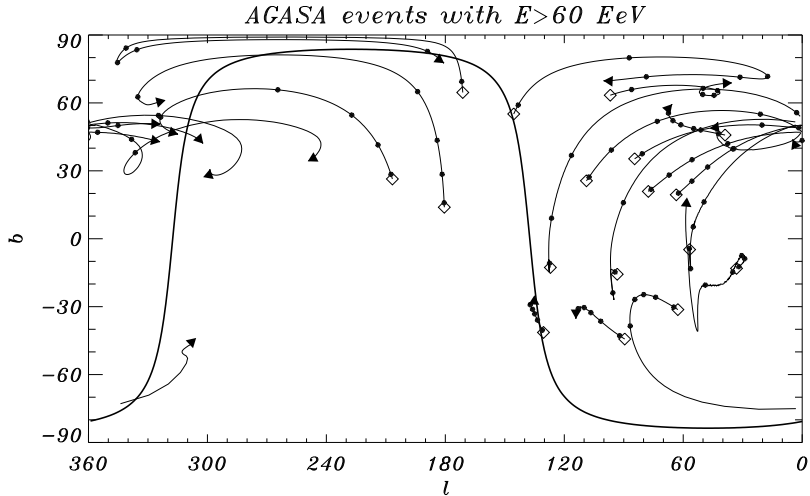} 
\caption{Observed arrival directions (diamonds) of AGASA events with
energies $>6\times 10^{19}$~eV and the corresponding incoming directions
outside the Galaxy for different CR charge $Z$. The dots along the lines
indicate the results for $Z=1,6,10,14$ and 20 and the tip of the arrow is
for $Z=26$ (iron). 
  \label{fig:agasa}}
\end{figure}

As we have shown \cite{ha99,ha00} magnetic deflections produce other
important effects which are even more striking. Indeed, the galactic
magnetic field acts as a giant lens and can magnify sizeably the CR fluxes
coming from any given source. Since the deflections are energy dependent,
this lensing effect will modify the original spectrum of the source.
Furthermore, since the B fields are not homogeneous, CRs from one source
may arrive, for a given energy, through more than one path, i.e. multiple
images of a source can be seen (fig.~3). The new images appear in pairs
(of
opposite parity) along critical lines in the sky seen on Earth,
corresponding to caustic lines in the source sky. When the source is on a
caustic, the magnification of the new pair of images is divergent, but for
decreasing energies the caustic moves away from the source and the
magnification behaves as 
$\mu_i\simeq A/\sqrt{1-E/E_0}\pm B+C_i\sqrt{1-E/E_0}$ near the energy
$E_0$ at which the
pair of images appeared \cite{ha00} (see fig.~4). 

\begin{figure}[t]
\epsfxsize=30pc 
\epsfbox{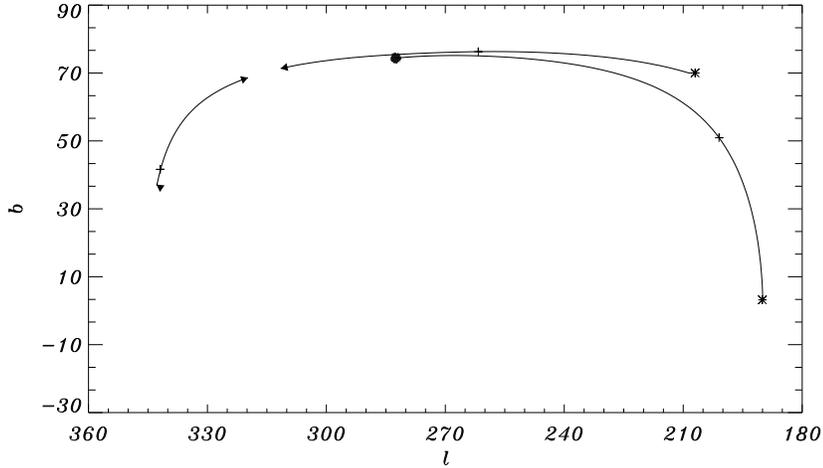} 
\caption{Illustration of the angular displacement of the image of a source
with energy and of the formation of secondary images in pairs. The actual
source (big dot) is in the direction of M87. The secondary images appear
where the two arrows meet. The images are located at the plus signs for
$E/Z=10^{19}$~eV  and at the asteriscs for $E/Z=5\times 10^{18}$~eV.
  \label{fig:images}}
\end{figure}

When convoluted with a
continuous
energy spectrum the divergence at the caustic is smoothed out, but anyhow
the large magnifications achieved make it more likely to detect events
at those energies. This may be helpful to account for some of
the doublets and triplets which have been observed, which actually tend to
be very close in energy as would be expected from clustering near a
caustic.

\begin{figure}[t]
\epsfxsize=28pc 
\epsfbox{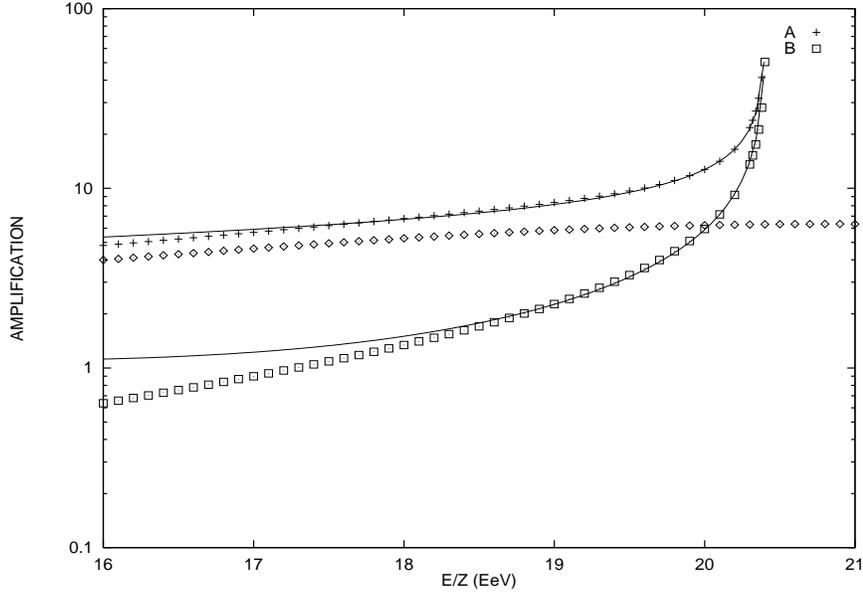} 
\caption{Numerical results and analytic fits to the magnification of the
CR flux near a caustic, where two images appear, together with the
original image (diamonds), for a source located in the direction of M87 in
the Virgo Cluster.
  \label{fig:spectrum}}
\end{figure}

There are many analogies\cite{ha99,ha00}. 
 between the features of magnetic lensing and the
more established gravitational lensing effect. This last is of course
achromatic, so that instead of changing the energy as in our discussion
above the analog would be to displace the source. 

The lensing effect can also modify the average composition arriving to the
Earth, since for a given energy the magnification of the fluxes depends on
the charge of the CR nuclei. Also significant time delays result from the
deflections which can be relevant in the observation of bursting sources.

All these effects are important if the ultra high energy CR sky is
dominated by a few powerful sources, as would be expected in AGN models.
If the CR flux were instead approximately isotropic (as happens at lower
energies), the Liouville theorem\cite{le33} would preclude the
observation of any
lensing effect: when a region of the sky is magnified, it is also seen
through a larger solid angle and the flux per unit solid angle remains
constant. Remarkably, the transition from one regime to the other seems to
be 
precisely around the ankle of the CR spectrum so that a host of
interesting effects may be studied with the expected increase in
statistics at the end of the CR spectrum.

 \section*{Acknowledgments}
Work partially supported by ANPCyT, CONICET and Fundaci\'on Antorchas,
Argentina.


\begin{thebibliography}{99}

\bibitem{gzk} K. Greisen, {\it Phys. Rev. Lett.} {\bf 16} (1996) 748; G.
T. Zatsepin and V. A. Kuzmin, {\it Sov. Phys. JETP} {\bf 4} (1966) 78.

\bibitem{pu76} J. L. Puget, F. W. Stecker and J. J. Bredekamp, {\it
Astrophys. J.} {\bf 205} (1976) 638; L. N. Epele and E. Roulet, {\it Phys.
Rev. Lett. }{\bf 81} (1999) 3295.

\bibitem{st97} T. Stanev, {\it Astrophys. J.} {\bf 479} (1997) 290; G.
Medina Tanco, E. Gouveia dal Pino and J. Horvath, {\it Astrophys. J. }{\bf
492} (1998) 200.


\bibitem{ha99} D. Harari, S. Mollerach and E. Roulet, {\it JHEP} {\bf 08} 
(1999) 022.

\bibitem{bh99} P. Bhattacharjee and G. Sigl, Phys. Rep. in press,
astro-ph/9811011.

\bibitem{ha00} D. Harari, S. Mollerach and E. Roulet, astro-ph/0001084.  

\bibitem{le33} G. Lema\^\i tre and M. S. Vallarta, {\it Phys.
Rev.} {\bf 44} (1933) 224.
\end{thebibliography}
\end{document}